\documentclass[aps,prb,twocolumn,showpacs,amsmath,amssymb,superscriptaddress]{revtex4-1}
\usepackage{graphicx}
\usepackage{dcolumn}
\usepackage{bm}
\usepackage{amsmath}

\begin{document}

\title{
Nucleation mechanism for the direct graphite-to-diamond phase transition
}

\author{Rustam Z. Khaliullin}
\email{rustam@khaliullin.com}
\author{Hagai Eshet}
\affiliation{
Department of Chemistry and Applied Biosciences, ETH Z\"urich, USI Campus, via G. Buffi 13, 6900 Lugano, Switzerland
}
\author{Thomas D. K\"uhne}
\affiliation{
Institute of Physical Chemistry, Johannes Gutenberg University Mainz, D-55128 Mainz, Germany
}
\affiliation{
Center for Computational Sciences, Johannes Gutenberg University Mainz, D-55128 Mainz, Germany
}
\author{J\"org Behler}
\affiliation{
Lehrstuhl f\"ur Theoretische Chemie, Ruhr-Universit\"at Bochum, D-44780 Bochum, Germany
}
\author{Michele Parrinello}
\affiliation{
Department of Chemistry and Applied Biosciences, ETH Z\"urich, USI Campus, via G. Buffi 13, 6900 Lugano, Switzerland
}

\date{\today}

\maketitle

\textbf{Graphite and diamond have comparable free energies, yet forming diamond from graphite is far from easy. In the absence of a catalyst, pressures that are significantly higher than the equilibrium coexistence pressures are required to induce the graphite-to-diamond transition~\cite{a:bundy1963,a:bundykasper1967,a:bundyrev,a:ultrahard,a:kurdyumov-nucleation-2004,a:sumiya2006,a:ohfuji2009}. Furthermore, the formation of the metastable hexagonal polymorph of diamond instead of the more stable cubic diamond is favored at lower temperatures~\cite{a:bundykasper1967,a:kurdyumov-nucleation-2004,a:sumiya2006,a:ohfuji2009}. The concerted mechanism suggested in previous theoretical studies~\cite{a:fahy1,a:fahy2,a:tsuneyuki,a:scandolo1,a:zipoli} cannot explain these phenomena. Using an \emph{ab initio} quality neural-network potential~\cite{a:khalcoex} we performed a large-scale study of the graphite-to-diamond transition assuming that it occurs via nucleation. The nucleation mechanism accounts for the observed phenomenology and reveals its microscopic origins. We demonstrated that the large lattice distortions that accompany the formation of the diamond nuclei inhibit the phase transition at low pressure and direct it towards the hexagonal diamond phase at higher pressure. The nucleation mechanism proposed in this work is an important step towards a better understanding of structural transformations in a wide range of complex systems such as amorphous carbon and carbon nanomaterials.}

Static compression of hexagonal graphite (HG) results in the formation of metastable hexagonal diamond (HD) at temperatures around 1200--1700~K~\cite{a:bundykasper1967,a:kurdyumov-nucleation-2004,a:sumiya2006,a:ohfuji2009} and cubic diamond (CD) at higher temperatures~\cite{a:bundy1963,a:bundyrev,a:ultrahard,a:kurdyumov-nucleation-2004,a:ohfuji2009}. Although the transition pressure is sensitive to the nature of the graphite samples neither of the diamond phases has been observed to form below $\sim$12~GPa. This pressure is significantly higher than the graphite-diamond coexistence pressure approximated by the Berman-Simon line $P(\text{GPa}) \sim 0.76 + 2.78 \times 10^{-3} T(\text{K})$~\cite{a:bundy1961}. 

Despite being an area of intense theoretical research~\cite{a:fahy1,a:fahy2,a:tsuneyuki,a:scandolo1,a:zipoli} the microscopic mechanism of the formation of metastable HD and the reason for the remarkable stability of graphite above the coexistence pressure are still unknown. Computer simulations, which could help resolve these issues, have been hindered because of the inability of empirical potentials to describe the energetics of the transformation accurately~\cite{a:khalcoex,a:gpp} and the computational expense of more reliable \emph{ab initio} methods. In the latter case, short simulation time and small system size (i.e. several hundred atoms) force the transition to occur in a concerted manner with the ultrafast ($\sim10^{-2}-1$~ps) synchronous formation of all new chemical bonds accross the entire simulation box~\cite{a:scandolo1,a:zipoli,a:shocktheor}. While concerted mechanisms can be observed at shock compression~\cite{a:shockexp1,a:shockexp2,a:shocktheor}, the transformation under static conditions is expected to proceed via nucleation and growth.

It has been estimated that because diamond has an extremely high surface energy~\cite{a:diamsur1} its critical nuclei may contain thousands of atoms~\cite{a:zerilli,a:diamnuc1,a:diamnuc2}. Hence, tens or even hundreds of thousands of atoms are required for modeling the diamond nuclei and the surrounding graphite matrix. Direct \emph{ab initio} simulations of systems of this size are outright impossible. Therefore, theoretical studies of the nucleation have been restricted to simple continuum models~\cite{a:zerilli,a:diamnuc1,a:diamnuc2}, which neglect the anisotropic nature of graphite, use significantly different estimates of the surface energy terms, and ignore the distortion of graphite around the growing nuclei.

In recent works, we have demonstrated that high-dimensional neural networks (NN)~\cite{a:behler} are capable of creating accurate representations of \emph{ab initio} potential energy surfaces of numerous elements~\cite{a:behler,a:khalcoex,a:khalNa1}. Even in the case of graphite and diamond which are very differently bonded, a NN potential predicts all relevant properties in quantitative agreement with \emph{ab initio} and experimental data (see Methods and Ref.~\onlinecite{a:khalcoex}). Furthermore, the computational efficiency of MD based on the NN potential enables us to extend time and length scales accessible to simulations and, thus, to perform the first atomistic study of homogeneous diamond nucleation from graphite.

\begin{figure*}
\includegraphics*[width=14.5cm]{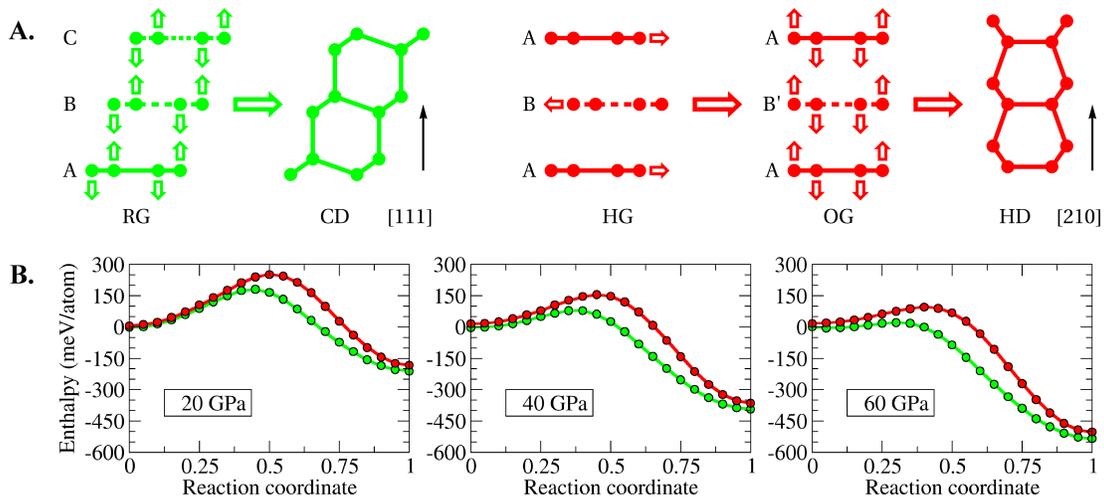}
\caption{\label{fig:nndft} A. Pathways for RG$\rightarrow$CD (green) and HG$\rightarrow$HD (red) transformations. OG denotes an (unstable) intermediate orthorhombic graphite phase. B. DFT (lines) and NN (circles) enthalpy profiles for the concerted mechanism: RG$\rightarrow$CD (green) and OG$\rightarrow$HD (red). The enthalpy of undistorted RG is taken as zero.
}
\end{figure*}

The energetics of the nucleation was studied at zero temperature by seeding diamond nuclei of various sizes inside a periodic $\sim$100$\times$100$\times$100~\AA\  graphite matrix containing $\sim$145,000 atoms (see Methods). Hexagonal and rhombohedral graphite (RG) lattices were used as initial structures for the formation of HD and CD nuclei, respectively, because of high symmetry of the HG$\rightarrow$HD and RG$\rightarrow$CD transformation pathways~\cite{a:fahy1,a:tsuneyuki}. HD and CD nuclei were formed by shifting atoms in the graphite lattice in the manner shown in Fig.~\ref{fig:nndft}A. Buckling of basal planes into the "chair" conformation in the RG lattice leads to CD whereas the "boat" buckling of the HG lattice results in the formation of HD. Such distortions of graphite planes have been observed in previous \emph{ab initio} simulations~\cite{a:scandolo1} and satisfy the orientation relations between graphite and diamond crystals discovered experimentally~\cite{a:kurdyumov-nucleation-2004,a:bundykasper1967,a:ohfuji2009,a:yagi-prb-hexd-1992}:
\begin{eqnarray}
\label{eq:orient}
\left( 100\right)_{HD} &\parallel &\left( 001\right)_{G} \text{ and } \left[ 100\right]_{HD} \parallel  \left[ 100\right]_{G},\nonumber\\
\left( 111\right)_{CD} &\parallel &\left( 001\right)_{G} \text{ and } \left[ 110\right]_{CD} \parallel  \left[ 100\right]_{G}.\nonumber
\end{eqnarray}

\begin{figure*}
\includegraphics*[width=17.5cm]{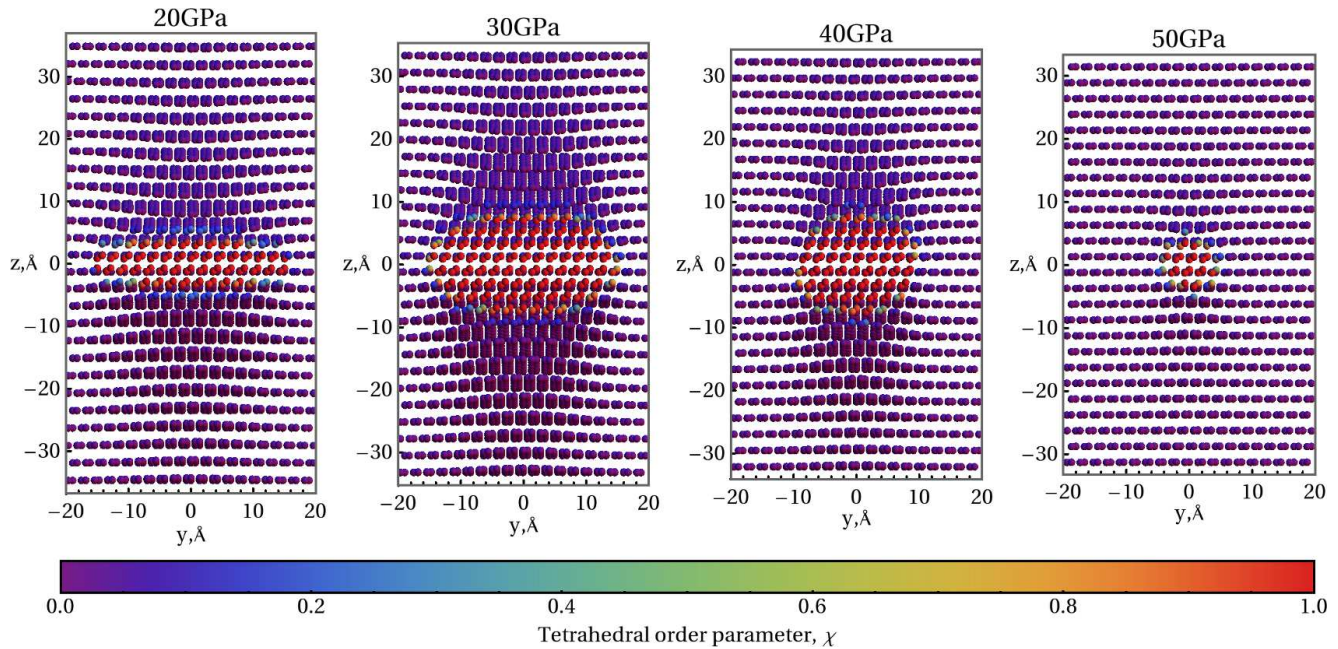}
\caption{\label{fig:cross} Pressure dependence of the shape and size of the CD nuclei. The optimized critical nuclei are shown for 30, 40, and 50~GPa whereas the 20~GPa nucleus is below the critical size and is presented to compare its shape with that of the 30~GPa nucleus. Atoms are colored according to the values of the tetrahedral atomic order parameter defined to distinguish graphite ($\chi_{i} \sim 0$) and diamond ($\chi_{i} \sim 1$) configurations (see Methods).}
\end{figure*}

\begin{figure}
\includegraphics*[width=8cm]{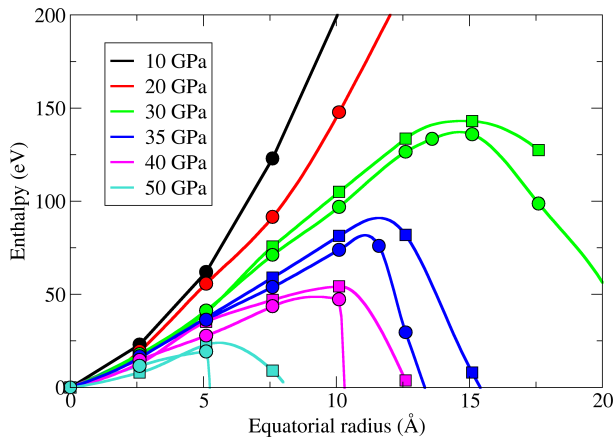}
\caption{\label{fig:barriers} Pressure dependence of the barriers for the RG$\rightarrow$CD (circles) and HG$\rightarrow$HD (squares) diamond nucleation.  The 20~GPa curve is predicted to reach the maximum around 25--31~\AA\ and 560--630~eV. The sharp drop of the curves at high pressure provides evidence that there are extremely low barriers for adding atoms to the critical nuclei that prevent stabilization of supercritical CD nuclei even at low temperatures.}
\end{figure}

Our calculations show that the nucleation process is strongly influenced by pressure. Both the nucleation barrier and the size of the critical nuclei decrease rapidly as the pressure is increased (Fig.~\ref{fig:cross} and~\ref{fig:barriers}). Examination of the atomic structure of the diamond nuclei reveals the microscopic origin of this phenomenon. Diamond nuclei are generally highly non-spherical and contain a diamond core (red atoms in Fig.~\ref{fig:cross}) separated from a relaxation region of graphite lattice by a thin ($\sim$5~\AA) high-enthalpy interface. To form interlayer bonds in the core, large regions in a number of graphite layers have to be buckled and bent in the $\left[ 001\right]_{G}$ direction. At low pressure these distortions are energetically costly because of the significant mismatch between the $\left[ 001\right]_{G}$ lattice parameters of the parent phase and the nuclei. At 10~GPa, we did not observe the formation of interlayer bonds even around large diamond seeds. At higher pressure the mismatch decreases and the diamond cores become increasingly more spherical (Fig.~\ref{fig:cross}). The same mismatch is the reason for the $\left[ 001\right]_{G}$ distortion of the graphite crystal around the nuclei. The extent of the distorted region decreases with the increase in pressure.

Since the size of the distorted region around the diamond core grows linearly with the size of the core the free energy of the diamond nucleus in graphite can be written as~\cite{a:olson,a:eshelby}
%
\begin{equation}
\label{eq:classical}
\Delta G = (\Delta g_s + \Delta g_{\mu}) V + \sigma S,
\end{equation}
where $\Delta g_{\mu}$ is the difference between the free-energy densities of bulk diamond and graphite, $\Delta g_s$ is the positive misfit strain free-energy per nucleus volume and $\sigma$ is the interfacial free-energy density. $S$ and $V$ are the surface area and volume of the nucleus, respectively. 
Due to the considerable mismatch between the lattice parameters, the large strain energy term outweighs the relatively small bulk term at $P<10$~GPa and makes nucleation impossible (i.e. the volume term becomes positive). As a result the 10~GPa nucleation curve does not show any signs of flattening out even at $R=50$~\AA. In contrast, the 20~GPa curve is predicted to reach the maximum around 25--31~\AA\ and 560--630~eV. Hence, our results are consistent with the observed $\sim$12~GPa minimum pressure threshold for diamond formation. The negative bulk term $\Delta g_{\mu}$ increases in magnitude with pressure (Fig.~\ref{fig:nndft}B) while the positive strain term becomes smaller (i.e. note the extent of the relaxation region around the diamond core in Fig.~\ref{fig:cross}) leading to the observed decrease in the nucleation barriers (Fig.~\ref{fig:barriers}). 

Comparison of the transition barriers for the concerted pathways and the nucleation process (Table~\ref{tab:compare}) shows that the nucleation is energetically more favorable in the pressure range used in compression experiments. Hence, nucleation represents a more realistic mechanism for diamond formation. However, at higher pressure ($\sim 50$~GPa) the graphite crystal approaches the lattice instability point and the activation barriers for the continuous transformations are lower (Fig.~\ref{fig:nndft}B and Table~\ref{tab:compare}). In this high-pressure limit, diamond domains can appear spontaneously throughout the graphite matrix without the formation of a well-defined graphite-diamond interface.  For example, at 60--70~GPa, the formation of only several interlayer bonds in close proximity to each other is enough to initiate a rapid irreversible growth of a diamond crystal.

\begin{table}
\caption{\label{tab:compare} Comparison of the enthalpy barriers of the concerted pathways and nucleation.}
\begin{ruledtabular}
\begin{tabular}{lccc}
Pressure &Nucleation\footnotemark[1] &\multicolumn{2}{c}{Concerted (meV/atom)}\\
(GPa)    &(meV/atom) & CD\footnotemark[2] & HD\footnotemark[2]\\
\hline \hline
30& 70--90 & 130 & 185 \\
40& 40--60 & 80 & 140 \\
50&  110--280 & 50 & 93 \\
\end{tabular}
\end{ruledtabular}
\footnotetext[1]{Calculated by dividing the nucleation barrier by the number of atoms in the diamond core ($\chi_i > 0.8$) of the critical nucleus. This value gives the upper bound on the nucleation barrier.}
\footnotetext[2]{Pathways are shown in Fig.~\ref{fig:nndft}A with green (CD) and red (HD) colors.}
\end{table}

The activation barriers for the concerted mechanism are determined by the energetics of buckling of graphite basal planes into the "chair" or "boat" configurations. The considerably lower activation barrier along the "chair" buckling mode favors the formation of CD (Fig.~\ref{fig:nndft}B and Table~\ref{tab:compare}) while in experiments the metastable HD phase is often observed. This disagreement between the theoretical predictions and experiment is resolved if the nucleation mechanism of diamond formation is considered. Our calculations show that the nucleation barriers for the HG$\rightarrow$HD and RG$\rightarrow$CD processes are not very different (Fig.~\ref{fig:barriers}). The similar heights of activation barriers is the consequence of close energy of CD and HD nuclei and their surface energies~\footnote{The rate of nucleation is controlled by the thermodynamic nucleation barrier and the activation barrier for bringing atoms across the interface. In the case of diamond, the latter term is small compared to the thermodynamic barriers and does not affect the overall kinetics considerably provided that the growth of the critical nuclei occurs in an atom-by-atom manner. Hence, we consider only the thermodynamic nucleation barriers in this work}.

In the nucleation process, the graphite basal planes undergo local "in-layer" distortions that bring atoms into appropriate stacking sequence inside diamond nuclei. The slightly higher barriers along the pathway leading to the HD phase are a consequence of this local $\left[ 210\right]_{G}$ distortion. This distortion brings the neighboring layers from the hexagonal (AB) to orthorhombic (AB') stacking (Fig.~\ref{fig:nndft}A) inside the HD nuclei. The HG$\rightarrow$CD nucleation process is considerably more difficult than both the RG$\rightarrow$CD and HG$\rightarrow$HD nucleation. The strains necessary to distort the lattice from the AB stacking in HG to ABC stacking inside CD nuclei prevent stabilization of the CD nuclei inside the HG lattice in our simulations.

In-layer distortions have not been observed in any of the previous theoretical studies of this phase transition~\cite{a:scandolo1,a:zipoli,a:shocktheor} because in-layer stresses result in the artificial sliding of graphite layers in small-cell simulations. Our results suggest that the in-layer distortions in large-cell simulations determine the energetics of the transformation and the overall direction for the phase transition. The higher in-layer strains for the HG$\rightarrow$HD transformation relative to the HG$\rightarrow$CD transition explain why HD is formed from graphite, which mostly consists of the hexagonal form. On the other hand, our studies show that the formation of CD is favored in the RG lattice because no in-layer distortions are required for the RG$\rightarrow$CD process. This implies that CD nucleates in the HG samples around stacking faults (e.g. $\left[ AB\right]_{n}\left[ CA\right]_{m}$), which are common defects in the HG lattice. High temperature fluctuations can also activate the sliding of graphite basal planes in HG with the formation of the RG regions and, hence, favor the nucleation of the CD phase.

In conclusion, our study of the homogeneous diamond nucleation offers new insights into the atomistic mechanism for the direct graphite-to-diamond phase transition. We have demonstrated that the transformation does not occur below $\sim$10~GPa in the static compression experiments because of the prohibitively large strains accompanying the formation of diamond nuclei at low pressure. We have also shown that, at higher pressures, the nucleation mechanism is favored over the concerted transformation. Larger distortions of the graphite lattice around the CD nuclei compared to those around the HD nuclei explain the formation of the metastable HD phase. At yet higher pressures, the transition is continuous and proceeds without formation of a well-defined graphite-diamond interface.

A finite temperature MD study of the thermodynamics and kinetics for heterogeneous nucleation on defects in the graphite lattice will be carried out in the future. The computational and theoretical models presented here offer new opportunities for investigating of complex structural transformations in a wide range of carbon-based materials.

\textbf{Methods.} The NN potential was created to reproduce the Perdew-Burke-Ernzerhof (PBE) density functional~\cite{a:pbe} energies for both phases in the pressure range up to 100~GPa, as described in our previous paper~\cite{a:khalcoex}. The PBE functional in combination with a dispersion corrected atom centered pseudopotential~\cite{a:dcacp} accounts for the van der Waals interactions in graphite and describes experimental structural, elastic and vibrational properties of the diamond and graphite phases~\cite{a:khalcoex}. The NN fitting procedure introduces only a small error (RMSE of the independent test sets is 4.0~meV/atom) in addition to the numerical (convergence) error of the DFT calculations. Thus, the NN-potential is expected to describe all properties and processes in graphite and diamond with an accuracy comparable with that of the PBE functional. 

Fig.~\ref{fig:nndft}B shows that the energetics of concerted transformations in small cells obtained with the NN potential is indistinguishable from the density functional theory (DFT) results. In quantitative agreement with the \emph{ab initio} calculations of Tateyama \emph{at al.}~\cite{a:tsuneyuki}, the stability of diamond relative to graphite increases with pressure whereas the barrier separating two phases decreases. At a pressure of 80~GPa graphite reaches a lattice instability point (i.e. one of the barriers becomes zero) and undergoes an ultrafast transformation to diamond as was previously observed in \emph{ab initio} simulations by Scandolo \emph{et al}~\cite{a:scandolo1}. The NN-driven MD simulations at constant pressure and temperature correctly reproduce the mechanism of this concerted transformation and are in perfect agreement with the results of Scandolo \emph{et al}.

\begin{figure}
\includegraphics*[width=8.5cm]{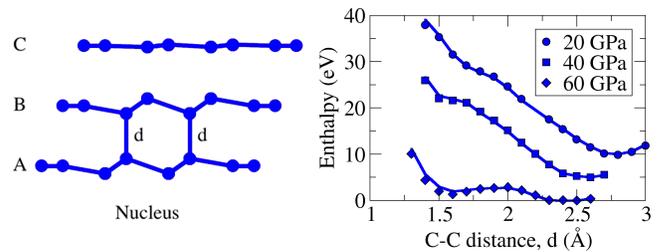}
\caption{\label{fig:nndft2} DFT (lines) and NN (circles) enthalpy profiles for a nucleation pathway of CD. The enthalpy of undistorted RG is taken as zero; the 20 and 40~GPa curves are shifted up for clarity of presentation.}
\end{figure}

Fig.~\ref{fig:nndft2} demonstrates that the formation of a diamond nucleus in a small system is also excellently described by the NN.

The energetics of the nucleation of diamond from graphite was studied at zero temperature on a model system of 145,152 carbon atoms arranged in a graphite lattice in a periodic $\sim$100$\times$100$\times$100~\AA\  simulation cell. Diamond nuclei of various shapes and sizes were seeded by shifting atoms within a certain predefined region of the cell in the directions shown in Fig.~\ref{fig:nndft}A and constraining distances between appropriate pairs of atoms to the values corresponding to the C--C distance in diamond. RG and HG was used as a starting lattice for the formation of CD and HD nuclei, respectively. 30~ps constant pressure MD simulations at 1000~K were performed to relax atoms around the constrained region followed by quenching to zero temperature. Finally, the constraints were removed and the geometry was optimized at constant pressure to obtain fully relaxed nuclei. We found that regardless of the shape of the initial constrained region all relaxed nuclei have shapes similar to those shown in Fig.~\ref{fig:cross}. We verified that all simulations results are converged with respect to the system size by performing additional test calculations for a $\sim$200$\times$200$\times$200~\AA\  simulation cell.

The tetrahedral order parameter $\chi_{i}$ for atom $i$ is defined as follows: $\chi_{i} = \frac{72}{n_i(n_i-1)} \sum_{j>k} n_{ij} n_{ik} (\frac{1}{2} + cos \theta_{jik})^{2}$, where the cutoff function is $n_{ij} = [1+e^{(r_{ij}-r_0)/\Delta r}]^{-1}$, the atom coordination number is $n_i = \sum_{j} n_{ij}$ and $r_0 = 1.6$~\AA, $\Delta r = 0.005$~\AA.

\textbf{Acknowledgments.} The authors would like to thank G. Tribello for carefully reading the manuscript and M. Ceriotti for useful discussions. JB is grateful for financial support by the FCI and the DFG. Our thanks are also due to the Swiss National Supercomputing Centre and High Performance Computing Group of ETH Z\"urich for computer time. 

\bibliography{carbon}

\begin{thebibliography}{30}%
\makeatletter
\providecommand \@ifxundefined [1]{%
 \@ifx{#1\undefined}
}%
\providecommand \@ifnum [1]{%
 \ifnum #1\expandafter \@firstoftwo
 \else \expandafter \@secondoftwo
 \fi
}%
\providecommand \@ifx [1]{%
 \ifx #1\expandafter \@firstoftwo
 \else \expandafter \@secondoftwo
 \fi
}%
\providecommand \natexlab [1]{#1}%
\providecommand \enquote  [1]{``#1''}%
\providecommand \bibnamefont  [1]{#1}%
\providecommand \bibfnamefont [1]{#1}%
\providecommand \citenamefont [1]{#1}%
\providecommand \href@noop [0]{\@secondoftwo}%
\providecommand \href [0]{\begingroup \@sanitize@url \@href}%
\providecommand \@href[1]{\@@startlink{#1}\@@href}%
\providecommand \@@href[1]{\endgroup#1\@@endlink}%
\providecommand \@sanitize@url [0]{\catcode `\\12\catcode `\$12\catcode
  `\&12\catcode `\#12\catcode `\^12\catcode `\_12\catcode `\%12\relax}%
\providecommand \@@startlink[1]{}%
\providecommand \@@endlink[0]{}%
\providecommand \url  [0]{\begingroup\@sanitize@url \@url }%
\providecommand \@url [1]{\endgroup\@href {#1}{\urlprefix }}%
\providecommand \urlprefix  [0]{URL }%
\providecommand \Eprint [0]{\href }%
\providecommand \doibase [0]{http://dx.doi.org/}%
\providecommand \selectlanguage [0]{\@gobble}%
\providecommand \bibinfo  [0]{\@secondoftwo}%
\providecommand \bibfield  [0]{\@secondoftwo}%
\providecommand \translation [1]{[#1]}%
\providecommand \BibitemOpen [0]{}%
\providecommand \bibitemStop [0]{}%
\providecommand \bibitemNoStop [0]{.\EOS\space}%
\providecommand \EOS [0]{\spacefactor3000\relax}%
\providecommand \BibitemShut  [1]{\csname bibitem#1\endcsname}%
\let\auto@bib@innerbib\@empty
\bibitem [{\citenamefont {Bundy}(1963)}]{a:bundy1963}%
  \BibitemOpen
  \bibfield  {author} {\bibinfo {author} {\bibfnamefont {F.~P.}\ \bibnamefont
  {Bundy}},\ }\href@noop {} {\bibfield  {journal} {\bibinfo  {journal} {J.
  Chem. Phys.}\ }\textbf {\bibinfo {volume} {38}},\ \bibinfo {pages} {631}
  (\bibinfo {year} {1963})}\BibitemShut {NoStop}%
\bibitem [{\citenamefont {Bundy}\ and\ \citenamefont
  {Kasper}(1967)}]{a:bundykasper1967}%
  \BibitemOpen
  \bibfield  {author} {\bibinfo {author} {\bibfnamefont {F.~P.}\ \bibnamefont
  {Bundy}}\ and\ \bibinfo {author} {\bibfnamefont {J.~S.}\ \bibnamefont
  {Kasper}},\ }\href@noop {} {\bibfield  {journal} {\bibinfo  {journal} {J.
  Chem. Phys.}\ }\textbf {\bibinfo {volume} {46}},\ \bibinfo {pages} {3437}
  (\bibinfo {year} {1967})}\BibitemShut {NoStop}%
\bibitem [{\citenamefont {Bundy}\ \emph {et~al.}(1996)\citenamefont {Bundy},
  \citenamefont {Bassett}, \citenamefont {Weathers}, \citenamefont {Hemley},
  \citenamefont {Mao},\ and\ \citenamefont {Goncharov}}]{a:bundyrev}%
  \BibitemOpen
  \bibfield  {author} {\bibinfo {author} {\bibfnamefont {F.~P.}\ \bibnamefont
  {Bundy}}, \bibinfo {author} {\bibfnamefont {W.~A.}\ \bibnamefont {Bassett}},
  \bibinfo {author} {\bibfnamefont {M.~S.}\ \bibnamefont {Weathers}}, \bibinfo
  {author} {\bibfnamefont {R.~J.}\ \bibnamefont {Hemley}}, \bibinfo {author}
  {\bibfnamefont {H.~K.}\ \bibnamefont {Mao}}, \ and\ \bibinfo {author}
  {\bibfnamefont {A.~F.}\ \bibnamefont {Goncharov}},\ }\href@noop {} {\bibfield
   {journal} {\bibinfo  {journal} {Carbon}\ }\textbf {\bibinfo {volume} {34}},\
  \bibinfo {pages} {141} (\bibinfo {year} {1996})}\BibitemShut {NoStop}%
\bibitem [{\citenamefont {Irifune}\ \emph {et~al.}(2003)\citenamefont
  {Irifune}, \citenamefont {Kurio}, \citenamefont {Sakamoto}, \citenamefont
  {Inoue},\ and\ \citenamefont {Sumiya}}]{a:ultrahard}%
  \BibitemOpen
  \bibfield  {author} {\bibinfo {author} {\bibfnamefont {T.}~\bibnamefont
  {Irifune}}, \bibinfo {author} {\bibfnamefont {A.}~\bibnamefont {Kurio}},
  \bibinfo {author} {\bibfnamefont {S.}~\bibnamefont {Sakamoto}}, \bibinfo
  {author} {\bibfnamefont {T.}~\bibnamefont {Inoue}}, \ and\ \bibinfo {author}
  {\bibfnamefont {H.}~\bibnamefont {Sumiya}},\ }\href@noop {} {\bibfield
  {journal} {\bibinfo  {journal} {Nature}\ }\textbf {\bibinfo {volume} {421}},\
  \bibinfo {pages} {806} (\bibinfo {year} {2003})}\BibitemShut {NoStop}%
\bibitem [{\citenamefont {Britun}\ \emph {et~al.}(2004)\citenamefont {Britun},
  \citenamefont {Kurdyumov},\ and\ \citenamefont
  {Petrusha}}]{a:kurdyumov-nucleation-2004}%
  \BibitemOpen
  \bibfield  {author} {\bibinfo {author} {\bibfnamefont {V.~F.}\ \bibnamefont
  {Britun}}, \bibinfo {author} {\bibfnamefont {A.~V.}\ \bibnamefont
  {Kurdyumov}}, \ and\ \bibinfo {author} {\bibfnamefont {I.~A.}\ \bibnamefont
  {Petrusha}},\ }\href@noop {} {\bibfield  {journal} {\bibinfo  {journal}
  {Powder Metall. Met. Ceram.}\ }\textbf {\bibinfo {volume} {43}},\ \bibinfo
  {pages} {87} (\bibinfo {year} {2004})}\BibitemShut {NoStop}%
\bibitem [{\citenamefont {Sumiya}\ \emph {et~al.}(2006)\citenamefont {Sumiya},
  \citenamefont {Yusa}, \citenamefont {Inoue}, \citenamefont {Ofuji},\ and\
  \citenamefont {Irifune}}]{a:sumiya2006}%
  \BibitemOpen
  \bibfield  {author} {\bibinfo {author} {\bibfnamefont {H.}~\bibnamefont
  {Sumiya}}, \bibinfo {author} {\bibfnamefont {H.}~\bibnamefont {Yusa}},
  \bibinfo {author} {\bibfnamefont {T.}~\bibnamefont {Inoue}}, \bibinfo
  {author} {\bibfnamefont {H.}~\bibnamefont {Ofuji}}, \ and\ \bibinfo {author}
  {\bibfnamefont {T.}~\bibnamefont {Irifune}},\ }\href@noop {} {\bibfield
  {journal} {\bibinfo  {journal} {High Pressure Res.}\ }\textbf {\bibinfo
  {volume} {26}},\ \bibinfo {pages} {63} (\bibinfo {year} {2006})}\BibitemShut
  {NoStop}%
\bibitem [{\citenamefont {Ohfuji}\ and\ \citenamefont
  {Kuroki}(2009)}]{a:ohfuji2009}%
  \BibitemOpen
  \bibfield  {author} {\bibinfo {author} {\bibfnamefont {H.}~\bibnamefont
  {Ohfuji}}\ and\ \bibinfo {author} {\bibfnamefont {K.}~\bibnamefont
  {Kuroki}},\ }\href@noop {} {\bibfield  {journal} {\bibinfo  {journal} {J.
  Mineral. Petrol. Sci.}\ }\textbf {\bibinfo {volume} {104}},\ \bibinfo {pages}
  {307} (\bibinfo {year} {2009})}\BibitemShut {NoStop}%
\bibitem [{\citenamefont {Fahy}\ \emph {et~al.}(1986)\citenamefont {Fahy},
  \citenamefont {Louie},\ and\ \citenamefont {Cohen}}]{a:fahy1}%
  \BibitemOpen
  \bibfield  {author} {\bibinfo {author} {\bibfnamefont {S.}~\bibnamefont
  {Fahy}}, \bibinfo {author} {\bibfnamefont {S.~G.}\ \bibnamefont {Louie}}, \
  and\ \bibinfo {author} {\bibfnamefont {M.~L.}\ \bibnamefont {Cohen}},\
  }\href@noop {} {\bibfield  {journal} {\bibinfo  {journal} {Phys. Rev. B}\
  }\textbf {\bibinfo {volume} {34}},\ \bibinfo {pages} {1191} (\bibinfo {year}
  {1986})}\BibitemShut {NoStop}%
\bibitem [{\citenamefont {Fahy}\ \emph {et~al.}(1987)\citenamefont {Fahy},
  \citenamefont {Louie},\ and\ \citenamefont {Cohen}}]{a:fahy2}%
  \BibitemOpen
  \bibfield  {author} {\bibinfo {author} {\bibfnamefont {S.}~\bibnamefont
  {Fahy}}, \bibinfo {author} {\bibfnamefont {S.~G.}\ \bibnamefont {Louie}}, \
  and\ \bibinfo {author} {\bibfnamefont {M.~L.}\ \bibnamefont {Cohen}},\
  }\href@noop {} {\bibfield  {journal} {\bibinfo  {journal} {Phys. Rev. B}\
  }\textbf {\bibinfo {volume} {35}},\ \bibinfo {pages} {7623} (\bibinfo {year}
  {1987})}\BibitemShut {NoStop}%
\bibitem [{\citenamefont {Tateyama}\ \emph {et~al.}(1996)\citenamefont
  {Tateyama}, \citenamefont {Ogitsu}, \citenamefont {Kusakabe},\ and\
  \citenamefont {Tsuneyuki}}]{a:tsuneyuki}%
  \BibitemOpen
  \bibfield  {author} {\bibinfo {author} {\bibfnamefont {Y.}~\bibnamefont
  {Tateyama}}, \bibinfo {author} {\bibfnamefont {T.}~\bibnamefont {Ogitsu}},
  \bibinfo {author} {\bibfnamefont {K.}~\bibnamefont {Kusakabe}}, \ and\
  \bibinfo {author} {\bibfnamefont {S.}~\bibnamefont {Tsuneyuki}},\ }\href@noop
  {} {\bibfield  {journal} {\bibinfo  {journal} {Phys. Rev. B}\ }\textbf
  {\bibinfo {volume} {54}},\ \bibinfo {pages} {14994} (\bibinfo {year}
  {1996})}\BibitemShut {NoStop}%
\bibitem [{\citenamefont {Scandolo}\ \emph {et~al.}(1995)\citenamefont
  {Scandolo}, \citenamefont {Bernasconi}, \citenamefont {Chiarotti},
  \citenamefont {Focher},\ and\ \citenamefont {Tosatti}}]{a:scandolo1}%
  \BibitemOpen
  \bibfield  {author} {\bibinfo {author} {\bibfnamefont {S.}~\bibnamefont
  {Scandolo}}, \bibinfo {author} {\bibfnamefont {M.}~\bibnamefont
  {Bernasconi}}, \bibinfo {author} {\bibfnamefont {G.~L.}\ \bibnamefont
  {Chiarotti}}, \bibinfo {author} {\bibfnamefont {P.}~\bibnamefont {Focher}}, \
  and\ \bibinfo {author} {\bibfnamefont {E.}~\bibnamefont {Tosatti}},\
  }\href@noop {} {\bibfield  {journal} {\bibinfo  {journal} {Phys. Rev. Lett.}\
  }\textbf {\bibinfo {volume} {74}},\ \bibinfo {pages} {4015} (\bibinfo {year}
  {1995})}\BibitemShut {NoStop}%
\bibitem [{\citenamefont {Zipoli}\ \emph {et~al.}(2004)\citenamefont {Zipoli},
  \citenamefont {Bernasconi},\ and\ \citenamefont {Martonak}}]{a:zipoli}%
  \BibitemOpen
  \bibfield  {author} {\bibinfo {author} {\bibfnamefont {F.}~\bibnamefont
  {Zipoli}}, \bibinfo {author} {\bibfnamefont {M.}~\bibnamefont {Bernasconi}},
  \ and\ \bibinfo {author} {\bibfnamefont {R.}~\bibnamefont {Martonak}},\
  }\href@noop {} {\bibfield  {journal} {\bibinfo  {journal} {Eur. Phys. J. B}\
  }\textbf {\bibinfo {volume} {39}},\ \bibinfo {pages} {41} (\bibinfo {year}
  {2004})}\BibitemShut {NoStop}%
\bibitem [{\citenamefont {Khaliullin}\ \emph {et~al.}(2010)\citenamefont
  {Khaliullin}, \citenamefont {Eshet}, \citenamefont {K\"uhne}, \citenamefont
  {Behler},\ and\ \citenamefont {Parrinello}}]{a:khalcoex}%
  \BibitemOpen
  \bibfield  {author} {\bibinfo {author} {\bibfnamefont {R.~Z.}\ \bibnamefont
  {Khaliullin}}, \bibinfo {author} {\bibfnamefont {H.}~\bibnamefont {Eshet}},
  \bibinfo {author} {\bibfnamefont {T.~D.}\ \bibnamefont {K\"uhne}}, \bibinfo
  {author} {\bibfnamefont {J.}~\bibnamefont {Behler}}, \ and\ \bibinfo {author}
  {\bibfnamefont {M.}~\bibnamefont {Parrinello}},\ }\href@noop {} {\bibfield
  {journal} {\bibinfo  {journal} {Phys. Rev. B}\ }\textbf {\bibinfo {volume}
  {81}},\ \bibinfo {pages} {100103} (\bibinfo {year} {2010})}\BibitemShut
  {NoStop}%
\bibitem [{\citenamefont {Bundy}\ \emph {et~al.}(1961)\citenamefont {Bundy},
  \citenamefont {Strong}, \citenamefont {Bovenkerk},\ and\ \citenamefont
  {Wentorf}}]{a:bundy1961}%
  \BibitemOpen
  \bibfield  {author} {\bibinfo {author} {\bibfnamefont {F.~P.}\ \bibnamefont
  {Bundy}}, \bibinfo {author} {\bibfnamefont {H.~M.}\ \bibnamefont {Strong}},
  \bibinfo {author} {\bibfnamefont {H.~P.}\ \bibnamefont {Bovenkerk}}, \ and\
  \bibinfo {author} {\bibfnamefont {R.~H.}\ \bibnamefont {Wentorf}},\
  }\href@noop {} {\bibfield  {journal} {\bibinfo  {journal} {J. Chem. Phys.}\
  }\textbf {\bibinfo {volume} {35}},\ \bibinfo {pages} {383} (\bibinfo {year}
  {1961})}\BibitemShut {NoStop}%
\bibitem [{\citenamefont {Bartok}\ \emph {et~al.}(2010)\citenamefont {Bartok},
  \citenamefont {Payne}, \citenamefont {Kondor},\ and\ \citenamefont
  {Csanyi}}]{a:gpp}%
  \BibitemOpen
  \bibfield  {author} {\bibinfo {author} {\bibfnamefont {A.~P.}\ \bibnamefont
  {Bartok}}, \bibinfo {author} {\bibfnamefont {M.~C.}\ \bibnamefont {Payne}},
  \bibinfo {author} {\bibfnamefont {R.}~\bibnamefont {Kondor}}, \ and\ \bibinfo
  {author} {\bibfnamefont {G.}~\bibnamefont {Csanyi}},\ }\href@noop {}
  {\bibfield  {journal} {\bibinfo  {journal} {Phys. Rev. Lett.}\ }\textbf
  {\bibinfo {volume} {104}},\ \bibinfo {pages} {136403} (\bibinfo {year}
  {2010})}\BibitemShut {NoStop}%
\bibitem [{\citenamefont {Mundy}\ \emph {et~al.}(2008)\citenamefont {Mundy},
  \citenamefont {Curioni}, \citenamefont {Goldman}, \citenamefont {Kuo},
  \citenamefont {Reed}, \citenamefont {Fried},\ and\ \citenamefont
  {Ianuzzi}}]{a:shocktheor}%
  \BibitemOpen
  \bibfield  {author} {\bibinfo {author} {\bibfnamefont {C.~J.}\ \bibnamefont
  {Mundy}}, \bibinfo {author} {\bibfnamefont {A.}~\bibnamefont {Curioni}},
  \bibinfo {author} {\bibfnamefont {N.}~\bibnamefont {Goldman}}, \bibinfo
  {author} {\bibfnamefont {I.~F.~W.}\ \bibnamefont {Kuo}}, \bibinfo {author}
  {\bibfnamefont {E.~J.}\ \bibnamefont {Reed}}, \bibinfo {author}
  {\bibfnamefont {L.~E.}\ \bibnamefont {Fried}}, \ and\ \bibinfo {author}
  {\bibfnamefont {M.}~\bibnamefont {Ianuzzi}},\ }\href@noop {} {\bibfield
  {journal} {\bibinfo  {journal} {J. Chem. Phys.}\ }\textbf {\bibinfo {volume}
  {128}},\ \bibinfo {pages} {184701} (\bibinfo {year} {2008})}\BibitemShut
  {NoStop}%
\bibitem [{\citenamefont {Erskine}\ and\ \citenamefont
  {Nellis}(1991)}]{a:shockexp1}%
  \BibitemOpen
  \bibfield  {author} {\bibinfo {author} {\bibfnamefont {D.~J.}\ \bibnamefont
  {Erskine}}\ and\ \bibinfo {author} {\bibfnamefont {W.~J.}\ \bibnamefont
  {Nellis}},\ }\href@noop {} {\bibfield  {journal} {\bibinfo  {journal}
  {Nature}\ }\textbf {\bibinfo {volume} {349}},\ \bibinfo {pages} {317}
  (\bibinfo {year} {1991})}\BibitemShut {NoStop}%
\bibitem [{\citenamefont {Erskine}\ and\ \citenamefont
  {Nellis}(1992)}]{a:shockexp2}%
  \BibitemOpen
  \bibfield  {author} {\bibinfo {author} {\bibfnamefont {D.~J.}\ \bibnamefont
  {Erskine}}\ and\ \bibinfo {author} {\bibfnamefont {W.~J.}\ \bibnamefont
  {Nellis}},\ }\href@noop {} {\bibfield  {journal} {\bibinfo  {journal} {J.
  Appl. Phys.}\ }\textbf {\bibinfo {volume} {71}},\ \bibinfo {pages} {4882}
  (\bibinfo {year} {1992})}\BibitemShut {NoStop}%
\bibitem [{\citenamefont {Vanderbilt}\ and\ \citenamefont
  {Louie}(1984)}]{a:diamsur1}%
  \BibitemOpen
  \bibfield  {author} {\bibinfo {author} {\bibfnamefont {D.}~\bibnamefont
  {Vanderbilt}}\ and\ \bibinfo {author} {\bibfnamefont {S.~G.}\ \bibnamefont
  {Louie}},\ }\href@noop {} {\bibfield  {journal} {\bibinfo  {journal} {Phys.
  Rev. B}\ }\textbf {\bibinfo {volume} {30}},\ \bibinfo {pages} {6118}
  (\bibinfo {year} {1984})}\BibitemShut {NoStop}%
\bibitem [{\citenamefont {Zerilli}\ and\ \citenamefont
  {Jones}(1996)}]{a:zerilli}%
  \BibitemOpen
  \bibfield  {author} {\bibinfo {author} {\bibfnamefont {F.~J.}\ \bibnamefont
  {Zerilli}}\ and\ \bibinfo {author} {\bibfnamefont {H.~D.}\ \bibnamefont
  {Jones}},\ }\href@noop {} {\bibfield  {journal} {\bibinfo  {journal} {AIP
  Conf. Proc.}\ }\textbf {\bibinfo {volume} {370}},\ \bibinfo {pages} {163}
  (\bibinfo {year} {1996})}\BibitemShut {NoStop}%
\bibitem [{\citenamefont {Bradley}(1971)}]{a:diamnuc1}%
  \BibitemOpen
  \bibfield  {author} {\bibinfo {author} {\bibfnamefont {R.~S.}\ \bibnamefont
  {Bradley}},\ }\href@noop {} {\bibfield  {journal} {\bibinfo  {journal} {J.
  Inorg. Nucl. Chem.}\ }\textbf {\bibinfo {volume} {33}},\ \bibinfo {pages}
  {1969} (\bibinfo {year} {1971})}\BibitemShut {NoStop}%
\bibitem [{\citenamefont {Deryagin}\ and\ \citenamefont
  {Fedoseev}(1979)}]{a:diamnuc2}%
  \BibitemOpen
  \bibfield  {author} {\bibinfo {author} {\bibfnamefont {B.~V.}\ \bibnamefont
  {Deryagin}}\ and\ \bibinfo {author} {\bibfnamefont {D.~V.}\ \bibnamefont
  {Fedoseev}},\ }\href@noop {} {\bibfield  {journal} {\bibinfo  {journal}
  {Bull. Acad. Sci. USSR Division Chem. Sci.}\ }\textbf {\bibinfo {volume}
  {28}},\ \bibinfo {pages} {1106} (\bibinfo {year} {1979})}\BibitemShut
  {NoStop}%
\bibitem [{\citenamefont {Behler}\ and\ \citenamefont
  {Parrinello}(2007)}]{a:behler}%
  \BibitemOpen
  \bibfield  {author} {\bibinfo {author} {\bibfnamefont {J.}~\bibnamefont
  {Behler}}\ and\ \bibinfo {author} {\bibfnamefont {M.}~\bibnamefont
  {Parrinello}},\ }\href@noop {} {\bibfield  {journal} {\bibinfo  {journal}
  {Phys. Rev. Lett.}\ }\textbf {\bibinfo {volume} {98}},\ \bibinfo {pages}
  {146401} (\bibinfo {year} {2007})}\BibitemShut {NoStop}%
\bibitem [{\citenamefont {Eshet}\ \emph {et~al.}(2010)\citenamefont {Eshet},
  \citenamefont {Khaliullin}, \citenamefont {K\"uhne}, \citenamefont {Behler},\
  and\ \citenamefont {Parrinello}}]{a:khalNa1}%
  \BibitemOpen
  \bibfield  {author} {\bibinfo {author} {\bibfnamefont {H.}~\bibnamefont
  {Eshet}}, \bibinfo {author} {\bibfnamefont {R.~Z.}\ \bibnamefont
  {Khaliullin}}, \bibinfo {author} {\bibfnamefont {T.~D.}\ \bibnamefont
  {K\"uhne}}, \bibinfo {author} {\bibfnamefont {J.}~\bibnamefont {Behler}}, \
  and\ \bibinfo {author} {\bibfnamefont {M.}~\bibnamefont {Parrinello}},\
  }\href@noop {} {\bibfield  {journal} {\bibinfo  {journal} {Phys. Rev. B}\
  }\textbf {\bibinfo {volume} {81}},\ \bibinfo {pages} {184107} (\bibinfo
  {year} {2010})}\BibitemShut {NoStop}%
\bibitem [{\citenamefont {Yagi}\ \emph {et~al.}(1992)\citenamefont {Yagi},
  \citenamefont {Utsumi}, \citenamefont {Yamakata}, \citenamefont {Kikegawa},\
  and\ \citenamefont {Shimomura}}]{a:yagi-prb-hexd-1992}%
  \BibitemOpen
  \bibfield  {author} {\bibinfo {author} {\bibfnamefont {T.}~\bibnamefont
  {Yagi}}, \bibinfo {author} {\bibfnamefont {W.}~\bibnamefont {Utsumi}},
  \bibinfo {author} {\bibfnamefont {M.}~\bibnamefont {Yamakata}}, \bibinfo
  {author} {\bibfnamefont {T.}~\bibnamefont {Kikegawa}}, \ and\ \bibinfo
  {author} {\bibfnamefont {O.}~\bibnamefont {Shimomura}},\ }\href@noop {}
  {\bibfield  {journal} {\bibinfo  {journal} {Phys. Rev. B}\ }\textbf {\bibinfo
  {volume} {46}},\ \bibinfo {pages} {6031} (\bibinfo {year}
  {1992})}\BibitemShut {NoStop}%
\bibitem [{\citenamefont {Chu}\ \emph {et~al.}(2000)\citenamefont {Chu},
  \citenamefont {Moran}, \citenamefont {Reid},\ and\ \citenamefont
  {Olson}}]{a:olson}%
  \BibitemOpen
  \bibfield  {author} {\bibinfo {author} {\bibfnamefont {Y.~A.}\ \bibnamefont
  {Chu}}, \bibinfo {author} {\bibfnamefont {B.}~\bibnamefont {Moran}}, \bibinfo
  {author} {\bibfnamefont {A.~C.~E.}\ \bibnamefont {Reid}}, \ and\ \bibinfo
  {author} {\bibfnamefont {G.~B.}\ \bibnamefont {Olson}},\ }\href@noop {}
  {\bibfield  {journal} {\bibinfo  {journal} {Metall. Mater. Trans. A}\
  }\textbf {\bibinfo {volume} {31}},\ \bibinfo {pages} {1321} (\bibinfo {year}
  {2000})}\BibitemShut {NoStop}%
\bibitem [{\citenamefont {Eshelby}(1957)}]{a:eshelby}%
  \BibitemOpen
  \bibfield  {author} {\bibinfo {author} {\bibfnamefont {J.~D.}\ \bibnamefont
  {Eshelby}},\ }\href@noop {} {\bibfield  {journal} {\bibinfo  {journal} {Proc.
  R. Soc. London, A}\ }\textbf {\bibinfo {volume} {241}},\ \bibinfo {pages}
  {376} (\bibinfo {year} {1957})}\BibitemShut {NoStop}%
\bibitem [{Note1()}]{Note1}%
  \BibitemOpen
  \bibinfo {note} {The rate of nucleation is controlled by the thermodynamic
  nucleation barrier and the activation barrier for bringing atoms across the
  interface. In the case of diamond, the latter term is small compared to the
  thermodynamic barriers and does not affect the overall kinetics considerably
  provided that the growth of the critical nuclei occurs in an atom-by-atom
  manner. Hence, we consider only the thermodynamic nucleation barriers in this
  work}\BibitemShut {NoStop}%
\bibitem [{\citenamefont {Perdew}\ \emph {et~al.}(1996)\citenamefont {Perdew},
  \citenamefont {Burke},\ and\ \citenamefont {Ernzerhof}}]{a:pbe}%
  \BibitemOpen
  \bibfield  {author} {\bibinfo {author} {\bibfnamefont {J.~P.}\ \bibnamefont
  {Perdew}}, \bibinfo {author} {\bibfnamefont {K.}~\bibnamefont {Burke}}, \
  and\ \bibinfo {author} {\bibfnamefont {M.}~\bibnamefont {Ernzerhof}},\
  }\href@noop {} {\bibfield  {journal} {\bibinfo  {journal} {Phys. Rev. Lett.}\
  }\textbf {\bibinfo {volume} {77}},\ \bibinfo {pages} {3865} (\bibinfo {year}
  {1996})}\BibitemShut {NoStop}%
\bibitem [{\citenamefont {von Lilienfeld}\ \emph {et~al.}(2004)\citenamefont
  {von Lilienfeld}, \citenamefont {Tavernelli}, \citenamefont {Rothlisberger},\
  and\ \citenamefont {Sebastiani}}]{a:dcacp}%
  \BibitemOpen
  \bibfield  {author} {\bibinfo {author} {\bibfnamefont {O.~A.}\ \bibnamefont
  {von Lilienfeld}}, \bibinfo {author} {\bibfnamefont {I.}~\bibnamefont
  {Tavernelli}}, \bibinfo {author} {\bibfnamefont {U.}~\bibnamefont
  {Rothlisberger}}, \ and\ \bibinfo {author} {\bibfnamefont {D.}~\bibnamefont
  {Sebastiani}},\ }\href@noop {} {\bibfield  {journal} {\bibinfo  {journal}
  {Phys. Rev. Lett.}\ }\textbf {\bibinfo {volume} {93}},\ \bibinfo {pages}
  {153004} (\bibinfo {year} {2004})}\BibitemShut {NoStop}%
\end{thebibliography}%

\end{document}